
%
%
%
\def\unredoffs{} \def\redoffs{\voffset=-.31truein\hoffset=-.59truein}
\def\speclscape{\special{ps: landscape}}
%
%
%
%
\newbox\leftpage \newdimen\fullhsize \newdimen\hstitle \newdimen\hsbody
\tolerance=1000\hfuzz=2pt
\catcode`\@=11 
\def\bigans{b }
\message{ big or little (b/l)? }\read-1 to\answ
\ifx\answ\bigans\message{(This will come out unreduced.}
\magnification=1200\unredoffs\baselineskip=24pt plus 2pt minus 1pt
\hsbody=\hsize \hstitle=\hsize 
\else\message{(This will be reduced.} \let\l@r=L
\magnification=1000\baselineskip=16pt plus 2pt minus 1pt \vsize=7truein
\redoffs \hstitle=8truein\hsbody=4.75truein\fullhsize=10truein\hsize=\hsbody
\output={\ifnum\pageno=0 
  \shipout\vbox{\speclscape{\hsize\fullhsize\makeheadline}
    \hbox to \fullhsize{\hfill\pagebody\hfill}}\advancepageno
  \else
  \almostshipout{\leftline{\vbox{\pagebody\makefootline}}}\advancepageno
  \fi}
\def\almostshipout#1{\if L\l@r \count1=1 \message{[\the\count0.\the\count1]}
      \global\setbox\leftpage=#1 \global\let\l@r=R
 \else \count1=2
  \shipout\vbox{\speclscape{\hsize\fullhsize\makeheadline}
      \hbox to\fullhsize{\box\leftpage\hfil#1}}  \global\let\l@r=L\fi}
\fi
%
\newcount\yearltd\yearltd=\year\advance\yearltd by -1900

\def\Title#1#2{\nopagenumbers\abstractfont\hsize=\hstitle\rightline{#1}%
\vskip 0.6in\centerline{\titlefont #2}\abstractfont\vskip .3in\pageno=0}
\def\Date#1{\vfill\leftline{#1}\tenpoint\supereject\global\hsize=\hsbody%
\footline={\hss\tenrm\folio\hss}}
%

\def\draftmode{\message{ DRAFTMODE }\def\draftdate{{\rm preliminary draft:
\number\month/\number\day/\number\yearltd\ \ \hourmin}}%
\headline={\hfil\draftdate}\writelabels\baselineskip=20pt plus 2pt minus 2pt
 {\count255=\time\divide\count255 by 60 \xdef\hourmin{\number\count255}
  \multiply\count255 by-60\advance\count255 by\time
  \xdef\hourmin{\hourmin:\ifnum\count255<10 0\fi\the\count255}}}
\def\nolabels{\def\wrlabeL##1{}\def\eqlabeL##1{}\def\reflabeL##1{}}
\def\writelabels{\def\wrlabeL##1{\leavevmode\vadjust{\rlap{\smash%
{\line{{\escapechar=` \hfill\rlap{\sevenrm\hskip.03in\string##1}}}}}}}%
\def\eqlabeL##1{{\escapechar-1\rlap{\sevenrm\hskip.05in\string##1}}}%
\def\reflabeL##1{\noexpand\llap{\noexpand\sevenrm\string\string\string##1}}}
\nolabels
%
\global\newcount\secno \global\secno=0
\global\newcount\meqno \global\meqno=1
\def\newsec#1{\global\advance\secno by1\message{(\the\secno. #1)}
\global\subsecno=0\eqnres@t\noindent{\bf\the\secno. #1}
\writetoca{{\secsym} {#1}}\par\nobreak\medskip\nobreak}
\def\eqnres@t{\xdef\secsym{\the\secno.}\global\meqno=1\bigbreak\bigskip}
\def\sequentialequations{\def\eqnres@t{\bigbreak}}\xdef\secsym{}
\global\newcount\subsecno \global\subsecno=0
\def\subsec#1{\global\advance\subsecno by1\message{(\secsym\the\subsecno. #1)}
\ifnum\lastpenalty>9000\else\bigbreak\fi
\noindent{\it\secsym\the\subsecno. #1}\writetoca{\string\quad
{\secsym\the\subsecno.} {#1}}\par\nobreak\medskip\nobreak}
\def\appendix#1#2{\global\meqno=1\global\subsecno=0\xdef\secsym{\hbox{#1.}}
\bigbreak\bigskip\noindent{\bf Appendix #1. #2}\message{(#1. #2)}
\writetoca{Appendix {#1.} {#2}}\par\nobreak\medskip\nobreak}
%
%
\def\eqnn#1{\xdef #1{(\secsym\the\meqno)}\writedef{#1\leftbracket#1}%
\global\advance\meqno by1\wrlabeL#1}
\def\eqna#1{\xdef #1##1{\hbox{$(\secsym\the\meqno##1)$}}
\writedef{#1\numbersign1\leftbracket#1{\numbersign1}}%
\global\advance\meqno by1\wrlabeL{#1$\{\}$}}
\def\eqn#1#2{\xdef #1{(\secsym\the\meqno)}\writedef{#1\leftbracket#1}%
\global\advance\meqno by1$$#2\eqno#1\eqlabeL#1$$}
%
\newskip\footskip\footskip14pt plus 1pt minus 1pt 
\def\footnotefont{\ninepoint}\def\f@t#1{\footnotefont #1\@foot}
\def\f@@t{\baselineskip\footskip\bgroup\footnotefont\aftergroup\@foot\let\next}
\setbox\strutbox=\hbox{\vrule height9.5pt depth4.5pt width0pt}
\global\newcount\ftno \global\ftno=0
\def\foot{\global\advance\ftno by1\footnote{$^{\the\ftno}$}}
%
\newwrite\ftfile
\def\footend{\def\foot{\global\advance\ftno by1\chardef\wfile=\ftfile
$^{\the\ftno}$\ifnum\ftno=1\immediate\openout\ftfile=foots.tmp\fi%
\immediate\write\ftfile{\noexpand\smallskip%
\noexpand\item{f\the\ftno:\ }\pctsign}\findarg}%
\def\footatend{\vfill\eject\immediate\closeout\ftfile{\parindent=20pt
\centerline{\bf Footnotes}\nobreak\bigskip\input foots.tmp }}}
\def\footatend{}
%
%
\global\newcount\refno \global\refno=1
\newwrite\rfile
\def\ref{[\the\refno]\nref}
\def\nref#1{\xdef#1{[\the\refno]}\writedef{#1\leftbracket#1}%
\ifnum\refno=1\immediate\openout\rfile=refs.tmp\fi
\global\advance\refno by1\chardef\wfile=\rfile\immediate
\write\rfile{\noexpand\item{#1\ }\reflabeL{#1\hskip.31in}\pctsign}\findarg}
\def\findarg#1#{\begingroup\obeylines\newlinechar=`\^^M\pass@rg}
{\obeylines\gdef\pass@rg#1{\writ@line\relax #1^^M\hbox{}^^M}%
\gdef\writ@line#1^^M{\expandafter\toks0\expandafter{\striprel@x #1}%
\edef\next{\the\toks0}\ifx\next\em@rk\let\next=\endgroup\else\ifx\next\empty%
\else\immediate\write\wfile{\the\toks0}\fi\let\next=\writ@line\fi\next\relax}}
\def\striprel@x#1{} \def\em@rk{\hbox{}}
\def\lref{\begingroup\obeylines\lr@f}
\def\lr@f#1#2{\gdef#1{\ref#1{#2}}\endgroup\unskip}

\def\addref#1{\immediate\write\rfile{\noexpand\item{}#1}} 
\def\startrefs#1{\immediate\openout\rfile=refs.tmp\refno=#1}
\def\xref{\expandafter\xr@f}\def\xr@f[#1]{#1}
\def\refs#1{\count255=1[\r@fs #1{\hbox{}}]}
\def\r@fs#1{\ifx\und@fined#1\message{reflabel \string#1 is undefined.}%
\nref#1{need to supply reference \string#1.}\fi%
\vphantom{\hphantom{#1}}\edef\next{#1}\ifx\next\em@rk\def\next{}%
\else\ifx\next#1\ifodd\count255\relax\xref#1\count255=0\fi%
\else#1\count255=1\fi\let\next=\r@fs\fi\next}
%

%
\newwrite\ffile\global\newcount\figno \global\figno=1
\def\fig{fig.~\the\figno\nfig}
\def\nfig#1{\xdef#1{fig.~\the\figno}%
\writedef{#1\leftbracket fig.\noexpand~\the\figno}%
\ifnum\figno=1\immediate\openout\ffile=figs.tmp\fi\chardef\wfile=\ffile%
\immediate\write\ffile{\noexpand\medskip\noexpand\item{Fig.\ \the\figno. }
\reflabeL{#1\hskip.55in}\pctsign}\global\advance\figno by1\findarg}
\def\xfig{\expandafter\xf@g}\def\xf@g fig.\penalty\@M\ {}
\def\figs#1{figs.~\f@gs #1{\hbox{}}}
\def\f@gs#1{\edef\next{#1}\ifx\next\em@rk\def\next{}\else
\ifx\next#1\xfig #1\else#1\fi\let\next=\f@gs\fi\next}
\newwrite\lfile
{\escapechar-1\xdef\pctsign{\string\%}\xdef\leftbracket{\string\{}
\xdef\rightbracket{\string\}}\xdef\numbersign{\string\#}}

\def\writestop{\def\writestoppt{\immediate\write\lfile{\string\pageno%
\the\pageno\string\startrefs\leftbracket\the\refno\rightbracket%
\string\def\string\secsym\leftbracket\secsym\rightbracket%
\string\secno\the\secno\string\meqno\the\meqno}\immediate\closeout\lfile}}
\def\writestoppt{}\def\writedef#1{}
\def\seclab#1{\xdef #1{\the\secno}\writedef{#1\leftbracket#1}\wrlabeL{#1=#1}}
\def\subseclab#1{\xdef #1{\secsym\the\subsecno}%
\writedef{#1\leftbracket#1}\wrlabeL{#1=#1}}
\newwrite\tfile \def\writetoca#1{}
\def\leaderfill{\leaders\hbox to 1em{\hss.\hss}\hfill}
\def\writetoc{\immediate\openout\tfile=toc.tmp
   \def\writetoca##1{{\edef\next{\write\tfile{\noindent ##1
   \string\leaderfill {\noexpand\number\pageno} \par}}\next}}}

%
\catcode`\@=12 
%
\edef\tfontsize{\ifx\answ\bigans scaled\magstep3\else scaled\magstep4\fi}
\font\titlerm=cmr7 \tfontsize \font\titlerms=cmr7 \tfontsize
\font\titlermss=cmr7 \tfontsize \font\titlei=cmmi10 \tfontsize
\font\titleis=cmmi7 \tfontsize \font\titleiss=cmmi5 \tfontsize
\font\titlesy=cmsy10 \tfontsize \font\titlesys=cmsy7 \tfontsize
\font\titlesyss=cmsy5 \tfontsize \font\titleit=cmti10 \tfontsize
\skewchar\titlei='177 \skewchar\titleis='177 \skewchar\titleiss='177
\skewchar\titlesy='60 \skewchar\titlesys='60 \skewchar\titlesyss='60
\def\titlefont{\def\rm{\fam0\titlerm}
\textfont0=\titlerm \scriptfont0=\titlerms \scriptscriptfont0=\titlermss
\textfont1=\titlei \scriptfont1=\titleis \scriptscriptfont1=\titleiss
\textfont2=\titlesy \scriptfont2=\titlesys \scriptscriptfont2=\titlesyss
\textfont\itfam=\titleit \def\it{\fam\itfam\titleit}\rm}
 \ifx\answ\bigans\else scaled\magstep1\fi
\ifx\answ\bigans\def\abstractfont{\tenpoint}\else
\font\abssl=cmsl10 scaled \magstep1
\font\absrm=cmr10 scaled\magstep1 \font\absrms=cmr7 scaled\magstep1
\font\absrmss=cmr5 scaled\magstep1 \font\absi=cmmi10 scaled\magstep1
\font\absis=cmmi7 scaled\magstep1 \font\absiss=cmmi5 scaled\magstep1
\font\abssy=cmsy10 scaled\magstep1 \font\abssys=cmsy7 scaled\magstep1
\font\abssyss=cmsy5 scaled\magstep1 \font\absbf=cmbx10 scaled\magstep1
\skewchar\absi='177 \skewchar\absis='177 \skewchar\absiss='177
\skewchar\abssy='60 \skewchar\abssys='60 \skewchar\abssyss='60
\def\abstractfont{\def\rm{\fam0\absrm}
\textfont0=\absrm \scriptfont0=\absrms \scriptscriptfont0=\absrmss
\textfont1=\absi \scriptfont1=\absis \scriptscriptfont1=\absiss
\textfont2=\abssy \scriptfont2=\abssys \scriptscriptfont2=\abssyss
\textfont\itfam=\bigit \def\it{\fam\itfam\bigit}\def\footnotefont{\tenpoint}%
\textfont\slfam=\abssl \def\sl{\fam\slfam\abssl}%
\textfont\bffam=\absbf \def\bf{\fam\bffam\absbf}\rm}\fi
\def\tenpoint{\def\rm{\fam0\tenrm}
\textfont0=\tenrm \scriptfont0=\sevenrm \scriptscriptfont0=\fiverm
\textfont1=\teni  \scriptfont1=\seveni  \scriptscriptfont1=\fivei
\textfont2=\tensy \scriptfont2=\sevensy \scriptscriptfont2=\fivesy
\textfont\itfam=\tenit \def\it{\fam\itfam\tenit}\def\footnotefont{\ninepoint}%
\textfont\bffam=\tenbf \def\bf{\fam\bffam\tenbf}\def\sl{\fam\slfam\tensl}\rm}
\font\ninerm=cmr9 \font\sixrm=cmr6 \font\ninei=cmmi9 \font\sixi=cmmi6
\font\ninesy=cmsy9 \font\sixsy=cmsy6 \font\ninebf=cmbx9
\font\nineit=cmti9 \font\ninesl=cmsl9 \skewchar\ninei='177
\skewchar\sixi='177 \skewchar\ninesy='60 \skewchar\sixsy='60
\def\ninepoint{\def\rm{\fam0\ninerm}
\textfont0=\ninerm \scriptfont0=\sixrm \scriptscriptfont0=\fiverm
\textfont1=\ninei \scriptfont1=\sixi \scriptscriptfont1=\fivei
\textfont2=\ninesy \scriptfont2=\sixsy \scriptscriptfont2=\fivesy
\textfont\itfam=\ninei \def\it{\fam\itfam\nineit}\def\sl{\fam\slfam\ninesl}%
\textfont\bffam=\ninebf \def\bf{\fam\bffam\ninebf}\rm}
%
%

\hyphenation{anom-aly anom-alies coun-ter-term coun-ter-terms}
\def\inv{^{\raise.15ex\hbox{${\scriptscriptstyle -}$}\kern-.05em 1}}

\def\Dsl{\,\raise.15ex\hbox{/}\mkern-13.5mu D} 
\def\dsl{\raise.15ex\hbox{/}\kern-.57em\partial}

\font\bigit=cmti10 scaled \magstep1
\def\lspace{\ifx\answ\bigans{}\else\qquad\fi}
\def\lbspace{\ifx\answ\bigans{}\else\hskip-.2in\fi} 
\def\boxeqn#1{\vcenter{\vbox{\hrule\hbox{\vrule\kern3pt\vbox{\kern3pt
	\hbox{${\displaystyle #1}$}\kern3pt}\kern3pt\vrule}\hrule}}}
\def\mbox#1#2{\vcenter{\hrule \hbox{\vrule height#2in
		\kern#1in \vrule} \hrule}}  
%

\def\darr#1{\raise1.5ex\hbox{$\leftrightarrow$}\mkern-16.5mu #1}

\def\roughly#1{\raise.3ex\hbox{$#1$\kern-.75em\lower1ex\hbox{$\sim$}}}

\rightline{ WISC-MILW-94-TH-26}

\Title{UCD-PHY-94-40}
  {{\vbox {\centerline{Phase Transition for  Gravitationally  Collapsing}\break
 \centerline{
Dust Shells  in  2+1 Dimensions}}}}

\centerline{Yoav Peleg}
\vskip .1in
\centerline{\it Department of Physics}
\centerline {\it   University of Wisconsin-Milwaukee}
\centerline{\it   Milwaukee, Wisconsin 53201 }
\centerline{\it yoav@alpha2.csd.uwm.edu}
\vskip .2in
\centerline{Alan R. Steif}
\vskip .1in
\centerline{\it Department of
Physics}
\centerline {\it   University of California }
  \centerline{\it Davis, CA 95616}
\centerline{\it steif@dirac.ucdavis.edu}
 \vskip .4in

\noindent
ABSTRACT: The collapse of thin dust shells in 2+1 dimensional gravity
with and without a cosmological constant is analyzed.  A critical value of the
shell's mass as a
function of its radius and position is derived. For $\Lambda <0$, a naked
singularity or black
hole forms depending on whether  the shell's mass is below or just above this
value.
The solution space is
divided into four different regions by three critical surfaces. For $\Lambda <
0$, two surfaces
separate regions of   black hole solutions and   solutions
with naked singularities,  while the other surface separates regions of
open and closed spaces.
Near the transition between black hole and naked singularity,
we find ${\cal M} \sim  c_{p}(p-p^{*})^{\beta}$,
where $\beta = 1/2$ and   ${\cal M}$ is a naturally defined order parameter.
   We find no phase transition in crossing from an  open to closed space.
The critical exponent appears to be universal for spherically symmetric
dust.
The critical solutions are analogous to higher dimensional
extremal black holes. All  four phases coexist at one point in solution
space corresponding to the static extremal solution.

\Date{ }
\def\L{\Lambda}
\def\m{\mu}
\def\n{\nu}
\def\rdot{ \dot r}
\def\p{\phi}
\def\s{\sigma}
\def\t{\tau}
\def\g{\gamma}
\def\ra{\rightarrow}
\def\3D{three\;dimensional}
{\bf Introduction:}
Following the work of Choptuik {\ref\Choptuik{ M.W. Choptuik,
{\it Phys. Rev. Lett.} {\bf 70} (1993) 9.}}, critical behavior has been
found in several models of black hole formation {\ref\Abr-Evan{A.M. Abrahams
and C.R. Evans, {\it Phys. Rev. Lett.} {\bf 70} (1993) 2980.}} \ref\Col-Evan{
{J. Coleman and  C.R. Evans, {\it Phys. Rev. Lett.} {\bf 72} (1994)
1842.}} \ref\stromthor{
{A. Strominger and L. Thorlacius, {\it Phys. Rev. Lett.} {\bf 72}
(1994) 1584.}} {\ref\Traschen{J. Traschen, University of Massachusetts
preprint, gr-qc/9403016.}} {\ref\husain{V. Husain, {\it et. al.},
University of Alberta preprint, gr-qc/9402021.}} {\ref\Brady{
{P.R. Brady, University
of Newcastle preprint NCL94-TP12, gr-qc/9409035.}}
\ref\Koik-Mish{{T. Koike and T. Mishima,
University of Tokyo preprint TIT/HEP-266, gr-qc/9409045.}}. In  these
models,  the space of solutions is separated by a critical surface into a
region of  black
hole solutions and a region of solutions which are not black holes. There exist
continuous
parameters with  critical values on the boundaries separating these regions.
The critical solutions are universal, and  scaling laws with critical exponents
have been found.
For the case of collapsing  spherically symmetric inhomogeneous dust,
the existence of different  phases
has been known for some time {\ref\Eard-Smarr{ D.M. Eardley and
L. Smarr, {\it Phys. Rev.} {\bf D19} (1979) 2239.}} {\ref\chris{D.
Christodoulou, {\it Comm. Math. Phys.} {\bf 93} (1984) 171.}}, and
recently the order
parameters and their critical values have been found {\ref\Joshi{P.S.
Joshi and T.P. Singh, TIFR preprint, gr-qc/9405036; I.H.
Dwivedi and P.S. Joshi,
  gr-qc/9405049; T.P. Singh and P.S. Joshi,
  gr-qc/9409062.}}.
So far,  the behavior near criticality has not been studied.

Since there is presently no deep understanding of this general phenomenon,
the study of simple models of gravitational collapse may be useful.
In this paper, we study collapsing spherical thin dust shells in $2+1$
dimensional gravity with both vanishing  and  negative cosmological constant.
Imposing junction conditions across the dust shell,
we derive a  relation for the  total mass of the system in terms of   the rest
mass of the dust shell $\m$,   its initial radius $r_{0}$, and
its initial velocity  $\dot{r}_{0}$.
We show that  the solution space is divided into four regions.
For  $\Lambda = 0$, the solutions in the four regions are: (1) open conical
spaces
{\ref\djt{S. Deser, R. Jackiw, and G.'t Hooft, {\it Ann. Phys.} (NY)
{\bf 152}  (1984) 220. }},
(2) open three-dimensional Misner-Taub-NUT (MTN) spacetimes,
(3) closed conical spaces {\djt} and (4) closed MTN spacetimes.
For  $\Lambda < 0$, the solutions in the four regions are:
(1) open anti-deSitter (adS) conical spaces {\ref\dj{S. Deser and R. Jackiw,
{\it Ann. Phys.} (NY) {\bf 153} (1984) 405.}},
(2) exteriors of three-dimensional black holes  {\ref\btz{ M. Banados, C.
Teitelboim, and
J. Zanelli, {\it Phys. Rev. Lett.} {\bf 69} (1992) 1849;  M. Banados, M.
Henneaux,  C. Teitelboim,
and J. Zanelli, {\it Phys. Rev. } {\bf D48} (1993) 1506.}},
(3) closed conical adS spaces {\dj} and (4) interiors of three-dimensional
black holes.
 We find the critical surfaces separating the four regions and study the
critical behavior
in their vicinity.  Following {\Choptuik-\Brady}, we define the order
parameter, ${\cal M}$,
as the total mass of the system. For $\Lambda < 0$, we show that near the two
critical
surfaces  separating  black holes and   solutions with naked singularities, the
order parameter
behaves as
\eqn\critic{
{\cal M} \sim c_{p} (p - p^{*})^{\beta}
,}
where $\beta = 1/2$ for both surfaces. $p$ is an affine parameter along
{\it any} curve in the space of solutions
that crosses the critical surface at $p=p^{*}$, and $c_{p}$ is a constant.
For $\Lambda =0$, there exists an analogous phase transition with the same
exponent. In going from  an open to closed space, we find
$\beta = 1$ suggesting that in that case   there is no phase transition.

\vskip 10pt
{\bf Junction Conditions :}
We now consider a collapsing spherical shell of dust in $2+1$ dimensions and
derive the junction
conditions which allow us to relate the exterior and interior geometries. The
trajectory of the
shell forms a two-dimensional hypersurface   $\Sigma$ in the $2+1$ dimensional
spacetime.
Let the exterior($+$) and interior($-$) spherically symmetric
metrics   be given by
\eqn\metric{
ds_{\pm}^2 = - A^{2}_{\pm} ( r_{\pm}) dt_{\pm}^2  + B^{2}_{\pm} ( r_{\pm})
dr_{\pm}^2 + r^2_{\pm} d \p^2 .}
The  surface stress tensor for dust is $S_{\m\n} = \s u_{\m} u_{\n}$ where
$u^{\m}
\equiv ({\partial \;\over \partial \tau})^{\m}$   and $\s$ are the 3-velocity
and the  mass
density of the shell with $\tau$ the proper time.  The junction conditions
across the shell
are 1) continuity of the induced two-dimensional metric $h_{ij}$ on $\Sigma$
and 2) a
discontinuity of the extrinsic curvature determined by the shell's stress
tensor. The components of
$  u^{\m}$ are $\dot t_{\pm},\,\rdot_{\pm}$ inside and outside the shell
where ${\dot{}} \equiv {d\;\over d\tau}$.
The induced $2$-metric is given by
\eqn\twometric{
dl^2 = -d\t^2 + r^2_{\pm}(\t) d\p^2 .
}
  Continuity of $h_{ij}$ across  the shell   yields
\eqn\requal{
r_+ (\t) = r_- (\t) \equiv R(\t)
.}
 The second junction condition can be obtained by  decomposing  Einstein's
equation into
components normal  and tangential to $\Sigma$. The components
of    the normal, $n^{\m}$, to the hypersurface  are $ n^t = \pm {B\over A}u^r$
and $ n^r = \mp {A\over B} u^t $.
Integrating the tangential components of  Einstein's equations across $\Sigma$,
one obtains the junction condition {\ref\isr{ W. Israel, {\it
Nuovo Cimento}, {\bf 44B}, (1966) 1.}}:
\eqn\junction{
[ K_{ij} - h_{ij} K^l_{\,l} ] = 8 \pi G S_{ij}
}
where  $K_{ij} \equiv h_i^k
h_j^l  \nabla_k n_l$  is   the extrinsic
curvature and  $i,j,...$ are tangential components.
The bracket denotes the discontinuity of the enclosed expression across  the
shell.
{}~From $[G_{ni}] =0$ and the junction condition {\junction}, one obtains the
conservation equation,
\eqn\conservation{
S_{i \;\;|j}^{\;j}  =0
}
where $|j$ is the covariant derivative in $\Sigma$.

The $\t \t$ component of {\junction} for the metric {\metric} becomes
\eqn\junc{
{1\over r} \left[ \left( {{1\over B^{2}}  + (u^r)^2}
\right)^{1/2} \right] = 8\pi G \s
.}
 Projecting the conservation equation {\conservation} onto $u^{\m}$
produces an equation for $\s$
 \eqn\projcons{
{d\s\over d\t} + u^i_{|i} \,\s =0
.}
  From the 2-metric {\twometric}, one finds  $u^i_{|i} = \rdot/r$.
Substituting into {\projcons} and solving for {$\sigma$} leads to
\eqn\sigm{
\s = {\m \over 2\pi r}
}
where the constant $\m$ is the rest mass of the shell. The other component of
{\conservation} implies that the dust trajectory is a geodesic
of the two-metric which  is already clear from the form of {\twometric}.
 Substituting {\sigm} into {\junc} yields
\eqn\master{
\hbox{sign}(n_{-})\left( {{1\over B_{-}^{2}}  + \rdot^2} \right)^{1/2} +
\hbox{sign}(n_{+})\left( {{1\over B_{+}^{2}}  + \rdot^2} \right)^{1/2} =
4G\mu
,}
where $\hbox{sign}(n_{-/+})$ is the orientation of the normal to $\Sigma$
in the inside/outside spaces respectively {\ref\GoldKatz{D.S. Goldwirth and
J. Katz, Tel-Aviv University and Hebrew University preprint,
gr-qc/9408034.}}.
One can easily see that the accelerations on both sides of the shell vanish.
This contrasts with $3+1$ dimensions where  the accelerations are non-zero, but
equal and
opposite on the two sides of the shell.

\vskip 10pt
{\bf $2+1$ Gravity ($\Lambda = 0$):}
All spacetime solutions to vacuum $2+1$ Einstein gravity with vanishing
cosmological constant
are locally flat, but can have non-trivial global identifications. Consider the
general static
spherically symmetric spacetime  {\djt}
\eqn\cone{
ds^2 = -\g dt^2 + {dr^2 \over \g}  + r^2 d\phi^2,\quad 0 \leq \phi \leq 2\pi
}
 where $\g$ is a constant. $\g=1$ corresponds to Minkowski space.
For $0 < \g \equiv \alpha^2 < 1$,  the spatial geometry of {\cone} describes a
cone
with a deficit angle
$\Delta \phi = (1 - \alpha)2\pi$  and  mass     $  m = ( 1 - \alpha )/4G$
{\djt}. For $\g<0$, {\cone} describes Taub-NUT (Misner) space
{\ref\HawEll{S.W. Hawking and
G.F.R. Ellis, {\it The large scale structure of space-time}, (Cambridge
University Press,
Cambridge, 1973).}}
with  $t$ spacelike and $r$ timelike.
Now, consider  a  dust shell with a flat interior geometry $(B_{-} = 1,
\hbox{sign}(n_{-})  =1 )$
and  with exterior geometry given by {\cone}. {\master} then yields
\eqn\flatjunc{
({ 1 + \rdot^2})^{1/2} + \hbox{sign}(n_{+})({ \g +
\rdot^2})^{1/2} = 4G \mu .
}
Since $\mu$ and $\g$ are constants, we recover the radial geodesic equation,
$\ddot{r} = 0$.

For a given $\mu$ and $\dot r$, {\flatjunc} determines the exterior geometry.
There is no
dependence on the radius of the shell because there is no length scale in 2+1
gravity with $\Lambda =0$. Consider
fixed $\rdot$ and increase $\m$ from zero. There are four ranges of $\m$
describing qualitatively
different exterior geometries shown in Fig. 1.

%

The four regions are:

I) $\quad 0<\m <\m_{c1}={1\over 4G} \left( (1+\rdot^2)^{1/2} - |\rdot |\right)$
\qquad $(\hbox{sign}(n_{+})  = -1 )$  \hfil \break
The exterior geometry is an open cone  with mass $m$ given by
\eqn\flatmass{
m = {1\over 4G} \biggl (
1 - \bigl ( { 1  - 8G\m ({1 +\rdot^2})^{1/2} +16G^{2}\mu^{2} }\bigr )^{1/2}
\biggr )
.}

At each time, $t$, space is described geometrically by a truncated cone
of deficit angle $\Delta \phi = 8\pi G m$.
The top of the cone corresponding to the inside of the shell is  replaced by a
flat disk (Fig. 2a).

For a shell at rest $(\rdot =0)$, $m=\mu$ is recovered.
As $\mu \rightarrow \mu_c = {1\over 4G}$, the cone limits to an infinite
cylinder   (Fig. 2b).
The coordinates $(t,r,\phi)$ are singular in that
case. One can define the coordinates
$T = \alpha t$ and $\rho = (r_{s} - r)/\alpha$ (where $r_{s}$ is the constant
radius of the shell). In the limit $\alpha \rightarrow 0$ we get the infinite
cylinder
metric $ds^{2} = -dT^{2} + d \rho^{2} + r_{s}^{2} d\phi^{2}$.
For $\rdot \neq 0$, the mass {\flatmass} depends on the kinetic energy of the
shell as well.
In this case, we can obtain the limiting geometry for $\mu\ra \m_{c1}$
by defining  the coordinates $  u = \alpha^2 t$ and
$  v = r/\alpha^2 - {|\dot{r}_{s}| \over \sqrt{\alpha^2 +
\dot{r}_{s}^{2} } } t$.\footnote{$\dagger$}{We would like to thank Jorma Louko
for
this important observation.} The metric then has a smooth $\alpha \rightarrow
0$ limit
approaching    $ds^{2}  =  -(\dot{r}_{s})^{-2} du^2 + 2 du dv + u^2
d \phi^2$. The vector $\partial / \partial \phi$ is
a null-rotation generator, and in
 Minkowskian coordinates $(x^{0},x^{1},x^{2})$, it can
indeed be expressed as a linear combination of rotation and boost generators
\eqn\nullrotation{
{\partial \over \partial \phi} = \left( x^{2} {\partial \over \partial x^{1}}
- x^{1} {\partial \over \partial x^{2}}  \right) +
\left( x^{2} {\partial \over \partial x^{0}} + x^{0} {\partial \over
\partial x^{2}} \right)
.}
The spacetime outside the shell is therefore
three-dimensional  Minkowski space with the null-rotation
identification $\phi \sim \phi + 2\pi$ {\ref\hs{G. Horowitz and A. Steif,  {\it
Phys. Lett. }
{\bf B258}  (1991) 91.}}.

II)  $\quad \m_{c1}<\m <\m_{c2}= {1\over 4G}(1+\rdot^2)^{1/2}  $ \qquad
$(\hbox{sign}(n_{+})  = -1 )$  \hfil \break
The exterior geometry is Taub-NUT, described by {\cone} with $\g < 0$ and $r >
r_{s}$.

III)  $\quad \m_{c2}<\m <\m_{c3}= {1\over 4G} \left( (1+\rdot^2)^{1/2} + |\rdot
|\right)$
\qquad $(\hbox{sign}(n_{+})  = +1 )$  \hfil \break
The exterior geometry is closed Taub-NUT, described by {\cone} with $\g < 0$
and $r < r_{s}$.

IV)  $\quad \m_{c3}<\m <\m_{c4}= {1\over 2G}(1+\rdot^2)^{1/2} $ \qquad
$(\hbox{sign}(n_{+})  = +1 )$  \hfil \break
The exterior geometry is  a closed cone (see fig. 2c) with mass
\eqn\closedmass{
m = {1\over 4G} \biggl( 1 + \bigl ( {1 -
8G\m({1 +\rdot^{2}})^{1/2} +16G^{2}\mu^2 }\bigr )^{1/2} \biggr )
.}
Now there is  an additional conical singularity outside the shell with mass
$m^{*} = {1 \over 2G} -m$ fixed by  the Euler number of the space (see fig.
2c).
As $\mu \ra \m_{c4}$, the geometry degenerates to a disk.
For $\mu > \mu_{c4}$, $m^{*}$ becomes negative so the physical solution space
is the region
$\mu < \mu_{c4}$ in Fig. 1.

{}~From Fig. 1, one observes  that $(\mu = 1/4G$, $\dot{r}=0)$ corresponding to
  the static
infinite cylinder  solution  (Fig. 2b)
is a special point in the
solution space, in which {\it all}   four phases coexist.
For a shell at rest $(\rdot =0)$ regions (II) and (III) are absent. As $\m $
exceeds $\mu_c = {1\over 4G}$,
the geometry goes directly from an open cone to a closed
cone with the infinite cylindrical geometry at $\mu=\mu_c$.
The mass $m_{c}$ is exactly the bound that was found in the general case
of 2+1 dimensional gravity {\ref\Ashtekar{A. Ashtekar and M. Varadarajan,
Penn St. preprint CGPG-94-6/3, gr-qc/9406040.}.
As was observed in {\djt}, one can go beyond $m = m_{c}$.
%

\vskip 10pt
{\bf $2+1$ Anti-deSitter Gravity ($\Lambda < 0$):}
All classical solutions to $2+1 $ gravity with
$\L<0$ correspond to anti-deSitter space (adS) locally. However,  non-trivial
global
identifications can lead to very different properties including different
values of the mass as
well as the existence of event horizons or naked singularities.
For the analysis of collapsing dust shells, it is most convenenient to describe
the solutions analytically.

The general spherically symmetric static solution in $2+1$ dimensional
cosmological gravity
can be written as
\eqn\djbtz{
ds^2 =  - ({r^2 \over l^2}+\g) dt^2     + ({r^2\over l^2} +
\g   )^{-1} dr^2 + r^2 d{\phi}^2,\quad  l = (- \Lambda)^{-1/2},\quad 0<\phi
< 2\pi
}
where $\g$ is a constant. $\g  = 1$ corresponds to  anti-deSitter (adS) space.
For $0< \g \equiv
\alpha^{2} < 1$, we have   anti-deSitter space with a deficit angle
$\Delta \phi = 2\pi (1 - \alpha)$   describing  a point particle with
a mass $m = (1-\alpha)/4G$ {\dj}. In the limit, $\Lambda \rightarrow 0$,
one recovers the point particle solutions in {\djt}.
For $\g < 0$ the solution {\djbtz} is a black hole with  ADM mass $M = - \g/8G
$
defined relative to the $M=0$ vacuum.
The singularity, $r=0$, is located behind
an event horizon at $r_H = \sqrt{8GM} l$. The black hole solutions approach
anti-deSitter space asymptotically. They  can also be obtained geometrically
from
anti-deSitter space by identifying under the action of a boost.
The $M=0$ solution is a non-singular spacetime
with  an infinite throat of zero radius.

One should distinguish the   BTZ mass $M$ (which is in fact the ADM
{\ref\ad{L. Abbott and
S. Deser, {\it Nucl. Phys.} {\bf B195} (1982) 76.}} mass)   from  the   DJ'tH
mass  $m$ defined
in {\djt  \dj}. The difference is due not only to different choices of
vacua, but to different choices of time slices as well.
The DJ'tH mass is defined as
{\djt} $m \equiv \int \sqrt{g^{(2)}} T^{0}_{0} d^{2} x$,
where $g^{(2)}$ is the {\it two}-dimensional metric on a spacelike surface.
On the other hand, the BTZ mass is defined as $M \equiv \int
\sqrt{-\bar{g}^{(3)}} T^{0}_{0}
d^{2} x$ where $\bar{g}^{(3)}$ is the {\it three}-dimensional
background metric.
In the definition of the  BTZ  mass,  $t$ in {\djbtz}
is the time parameter, while in the definition of the  DJ'tH mass,
it is  $\alpha t$.

Let us now consider the  collapse of a dust shell in the geometry {\djbtz}.
Let $ \g_{+}$ and $\g_-$ denote the values of $\g $
for the exterior and interior geometries.
  {\master} yields
\eqn\masterjunc{
\hbox{sign}(n_{-})({ \g_{-} + r^2/l^2+ \rdot^2})^{1/2} + \hbox{sign}(n_{+}) ({
\g_{+} +
r^2/l^2 + \rdot^2})^{1/2} = 4G\m .
}
 {\masterjunc} implies  $r^{2}/l^{2} + \dot{r}^{2}$ is a constant and is
consistent with   the radial geodesic equation $\ddot{r} = -{1\over l^{2}} r$.
  $\g_{+} > 0$ and $\g_{+} < 0$ correspond to the formation of  a conical naked
singularity
and  a black hole respectively.

For the case of  collapse into the anti-deSitter
vacuum $(\g_- =1,\; \hbox{sign}(n_{-})=1)$ with an exterior open space
 $(\g_+ = -8GM, \;\hbox{sign}(n_{+}) =-1) $, {\masterjunc}  yields
\eqn\adsmass{
M  =     -{1\over 8G} +({1+ r^2/l^2+\dot r^2})^{1/2} \mu   -2G\m^2
\quad\quad (\hbox{collapse in adS space}).
}
$M$ increases with  $\rdot$, the
 kinetic energy of the shell as well as with $r$
due to the increasing anti-deSitter potential energy. We see that $M$ can be
negative (naked
conical singularity) or positive (black hole), depending on $\m , r$ and $\dot
r$.
 For $M<0$, the corresponding DJ'tH mass is well-defined and is given by
\eqn\djmass{
m = {1\over 4G} \biggl ( 1 - \bigl ( 1 - 8G\m ( 1 + r^{2}/l^{2}
+ \dot r^{2} )^{1/2} + 16G^2\m^2 \bigr )^{1/2} \biggr )
\quad \quad(\hbox{collapse in adS space})
.}
For an exterior closed space ($\hbox{sign}(n_{+})  =1$) {\master} yields
\eqn\closedadsjunc{
({ \g_{-} + r^2/l^2+ \rdot^2})^{1/2} + ({ \g_{+} +
r^2/l^2 + \rdot^2})^{1/2} = 4G\m .
}
As a function of $\m$, $r$ and $\dot r$, the BTZ mass is  again
given by {\adsmass}  while the DJ'tH mass is now
\eqn\djclosed{
m = {1\over 4G} \biggl ( 1 + \bigl ({ 1 - 8G\m ({1 + r^{2}/l^{2}
+ \dot{r}^{2}})^{1/2} + 16G^2\m^2 \bigr )  }^{1/2} \biggr )
.}

The total mass {\adsmass} can be viewed as
a function of three parameters: the dust shell rest mass, $\m$, its initial
radius, $r_{0}=r(0)$ and its initial velocity, $ \dot{r}_{0} = \dot{r}(0)$.
The space of solutions is the three dimensional space ${\cal S} \equiv
\{ \mu, r_0, \dot{r}_{0} \}$, shown in fig. 3.
The solution space for  $\Lambda = 0$ (Fig. 1) is the restriction of  ${\cal
S}$ to the
$r_{0} = 0$  plane.

%
%
%

As in the $\L=0$ case, there are four qualitatively different regions in
${\cal S}$ corresponding
to four ranges of $\mu$ as a function of $r$ and $\rdot$.
The solutions in the four regions are:  \hfil \break
\indent $(I)$  $0 < \mu < \mu_{c1}$,  open conical adS spaces,  \hfil \break
\indent $(II)$  $\mu_{c1} < \mu < \mu_{c2}$,  exteriors of  BTZ black holes ($r
> r_{s}$),
\hfil \break
\indent $(III)$  $\mu_{c2} < \mu < \mu_{c3}$, interiors of  BTZ black holes
($r < r_{s}$),  \hfil \break
\indent $(IV)$ $\mu_{c3} < \mu < \mu_{c4}$, closed conical adS spaces. \hfil
\break
The critical values of $\mu$ separating the regions can be obtained from those
for
$\Lambda =0$ by the replacement $\dot r^2 \rightarrow \dot r^2 + r^2/l^2$.
On the surface separating closed and open spaces II and III, $\gamma =
-(r_{0}^{2}/l^2 + \dot{r}_{0}^2)$. The exterior geometries associated with
the solutions on the critical surface separating regions   I and II  (III and
IV) are the {\it exterior} ($r > r_{s}$) ({\it interior} ($r < r_{s}$)) of the
infinite throat  $M=0$ black hole solution.    The point ($\mu = 1/4G$,
$r_{0}=0$,
$\dot{r}_{0}=0$) is the {\it complete} infinite throat $M=0$ vacum solution.
This is
a special point in solution space in which {\it all} the four phases
coexist.\footnote{\ddag}{The
$M=0$ solution is very similar to higher dimensional
extremal black holes and its special role in the solution
space raises the question whether this is a general feature of extremal black
holes. }

Collapse in  the $M=0$ vacuum $(\g_-=0)$ results in one
of the following three possibilities: \hfil \break
1) $0 < \mu < {1\over 4G}(r^{2}/l^{2} + \dot{r}^{2})^{1/2}$, an open BTZ
black hole \hfil \break
2) ${1\over 4G}(r^{2}/l^{2} + \dot{r}^{2})^{1/2} < \mu <
{1\over 2G}(r^{2}/l^{2} + \dot{r}^{2})^{1/2}$, a closed BTZ black hole
\hfil \break
3) $ {1\over 2G}(r^{2}/l^{2} + \dot{r}^{2})^{1/2} < \mu$, a closed conical
adS space.

\vskip 10pt
{\bf Phase Transition and Critical Exponents:}
We  now study the structure of
the phase transition as one crosses the two $\g=0$ critical surfaces
separating regions (I) and (II) and regions (III) and (IV).
{}~Following {\Choptuik - \Brady},
we look for an order parameter related to    the total mass.
In approaching the critical surfaces from regions I and IV, we use the
following order
parameter which is related to the  DJ'tH mass and vanishes on both critical
surfaces
$(m_c = 1/4G)$:
\eqn\orderpar{
{\cal M}  \equiv \alpha
 = 1 - 4Gm .}
The   Minkowski space (for $\Lambda=0$) and  adS space (for $\Lambda < 0$)
solutions
with ${\cal M} = 1\; (m=0)$
might  be thought of as  the ordered vacua.

{}~From {\djmass} and {\djclosed}
one can obtain the behavior of ${\cal M}$ near the critical  surfaces.
In the direction of  the principle axes,
${\cal M}$ takes the form
\eqn\critical{
{\cal M} \sim  \cases{ \mp { (8G)^{1/2} ({(r^{\pm}_{0c})^{2}/l^{2}
+ (\dot{r}^{\pm}_{0c})^{2}})^{1/4} }  |\m - \m^{\pm}_{c}|^{1/2}
  & \cr
\mp {\left( {8G\m^{\pm}_c \dot{r}^{\pm}_{0c}\over 1 + (r^{\pm}_{0c})^{2}/l^{2}
+
(\dot{r}^{\pm}_{0c})^{2} } \right)}^{1/2}
|\dot{r}_{0} -  \dot{r}^{\pm}_{0c}|^{1/2}
  & \cr
\mp {\left( {8G\m^{\pm}_c r^{\pm}_{0c}l^{-2}\over 1 + (r^{\pm}_{0c})^{2}/l^{2}
+
(\dot{r}^{\pm}_{0c})^{2} } \right)}^{1/2} |r_{0} - r^{\pm}_{0c}|^{1/2}
}
}
where
$\mu^{\pm}_c (r^{\pm}_{0c},\dot{r}^{\pm}_{0c})$ can be obtained from {\adsmass}
by setting $M=0$.
The lower (upper) signs  in  {\critical} refer to the lower (upper) shaded
surfaces in Fig. 3.
{}~From {\critical},
we see that the critical exponent is $\beta =1/2$ as one approaches the
critical surface from {\it any} direction.

%
%

Consider approaching the critical surfaces from regions II and III.
The DJ'tH mass in those regions is ill-defined.
A geometrical quantity
which is sensitive to the phase transition and is a
continuation of $\alpha$ to regions II and III is the black hole radius,
$r_{H}$.\footnote{\S}{In
{\Choptuik - \Brady}, the black hole mass was used as order parameter, but in
four dimensions,
this is proportional to the black hole radius.}
Using {\adsmass}, we find  $r_{H}\sim c_{p} (p-p^{*})^{1/2}$  and
thus   again obtaining the same exponent $\beta =1/2$ as in {\critical}.

A value of $1/2$ for the critical exponent $\beta$ was found in certain models
in
$3+1$ dimensions {\Traschen}{\husain}{\Brady}
and also in $1+1$ dimensions{\stromthor}. Our critical
solution does  not appear to be self-similar. So it is not clear if there is a
deeper
relation between these models and the one considered here. However,
since our critical solution is very similar to a higher dimensional  extremal
black hole,
a natural question is whether a phase transition also
occurs   in other models of  the formation of extremal black holes.
Some indications for that can be found in {\Traschen}.

Near  the  $\g = -(r_{0}^{2}/l^{2} + \dot{r}_{0}^{2})$  surface,
one finds $\beta = 1$ using either  $\alpha$ or $r_{H}$ or $M$ as order
parameter. Thus, there is no phase transition in crossing from an open to
closed space. A way to
see the distinct behavior of the transition between  open and closed spaces is
to consider
the static   case with $\Lambda =0$. Now, the only parameter is $\mu$,
and  the transition   connects regions I and IV directly.
We find ${\cal M} =  ~  c_{\m} (\m - \m^{*})$,
and thus indeed $\beta = 1$.

\vskip 10pt
{\bf Summary :}
In this paper, we described the phase transition which occurs in the
gravitational collapse of
dust shells in 2+1 dimensional gravity.
We found that the solution space is divided into four qualitatively distinct
regions.
There is   critical behavior near the transition between  black holes and
solutions with  naked singularities.
One can easily generalize our results to higher (or lower) dimensional
solution spaces. Consider for example   collapse into a conical
space, with interior mass $m_{0}$ (Fig. 2d).  We  then have a four dimensional
solution space,
and furthermore, one finds  that ${\cal M} \sim c_{m_{0}} (m_{0} -
m_{0c})^{1/2}$   as well. On the other hand, in the case
of the collapse of a homogeneous ball of  dust  in 2+1 dimensions
{\ref\mr{R.B. Mann and S.F. Ross, {\it Phys. Rev.} {\bf D47} (1993) 3319.},
the total mass is given by $M = \mu - 1/8G$ and is therefore only  a function
of $\m$.
So we have effectively a one-dimensional solution space, and if we
use our definition for the order parameter, we find  ${\cal M} =
(\m - \tilde{\m}_{c} )^{1/2}$, where $\tilde{\m}_{c} = 1/8G$.
So, it seems that our results and in particular the value of the critical
exponent are quite general in the case of spherical dust collapse in 2+1
dimensions.

\vskip 20pt

\centerline{\bf Acknowledgements}

We would like to thank S.  Carlip, S. Deser, J. Louko, R. Mann and L. Parker
for very
helpful discussions. Y.P. would like to acknowledge the financial support
of the NSF grants NSF-91-05935 and NSF-93-15811.
A.S. would like to acknowledge the financial support of the
SERC at Cambridge and NSF grant NSF-PHY-93-57203 at Davis.

\baselineskip=30pt

\footatend\vfill\supereject\immediate\closeout\rfile\writestoppt
\baselineskip=14pt\centerline{{\bf References}}\bigskip{\frenchspacing%
\parindent=20pt\escapechar=` \input refs.tmp\vfill\eject}\nonfrenchspacing

{\titlefont Figures Captions}

\vskip 25pt

{\it Figure 1}: Solution space for $\L = 0$ with $G = 1/4$. The lower and upper
solid lines correspond to $\mu = \mu_{c1}$ ($\g = 0$) and $\m = \m_{c3}$
($\g=0$)
respectively. The dashed line corresponds to $\m = \m_{c2}$ ($\g =
-\dot{r}^{2}$).
The upper dotted line corresponds to $\m = \m_{c4}$. The solutions in the four
regions are (I) open cones, (II) open Taub-NUT spaces, (III) closed Taub-NUT
spaces,
and (IV) closed cones.

\vskip 15pt

{\it Figure 2}: Spatial 2-geometries for $\L = 0$ and $\dot{r} = 0$ : a) open
conical
space ($m < m_{c} = 1/4G$), b) the infinite cylinder ($m = m_{c}$), c) closed
conical
space ($m > m_{c}$), d) collapse onto a particle of mass $m_{0}$.

\vskip 15pt

{\it Figure 3}: Solution space ${\cal S}$ for shell collapse in adS vacuum
($G=1/4$).

\vskip 15pt

{\it Figure 4}: The order parameter, $\alpha$, as a function of $p =
\dot{r}_{0}$ with
$r^{+}_{0} / l = r^{-}_{0} / l = 1/4$ and $4G\m^{+} = (4G\m^{-})^{-1} =
2^{1/2}$.
The upper part, $\alpha > 0$, corresponds to region I in fig. 3, and the lower
part,
$\alpha < 0$, corresponds to region IV.

\end